\documentclass{aa}
\usepackage{epsfig,amsfonts,amssymb}
\begin{document}

\thesaurus{10.08.1, 10.11.1, 10.19.2, 12.04.1, 12.07.1}

\title{EROS 2 intensive observation of the caustic crossing 
of microlensing event MACHO SMC-98-1
\thanks{Based on observations made at the European Southern Observatory,
La Silla, Chile.}}
\author{
C.~Afonso\inst{1}, 
C.~Alard\inst{11},
J.N.~Albert\inst{2},
J.~Andersen\inst{6},
R.~Ansari\inst{2}, 
\'E.~Aubourg\inst{1}, 
P.~Bareyre\inst{1,4}, 
F.~Bauer\inst{1},
J.P.~Beaulieu\inst{5},
A.~Bouquet\inst{4},
S.~Char\inst{7},
X.~Charlot\inst{1},
F.~Couchot\inst{2}, 
C.~Coutures\inst{1}, 
F.~Derue\inst{2}, 
R.~Ferlet\inst{5},
J.F.~Glicenstein\inst{1},
B.~Goldman\inst{1,9,10},
A.~Gould\inst{1,8}\thanks{Alfred P.\ Sloan Foundation Fellow},
D.~Graff\inst{1,8},
M.~Gros\inst{1}, 
J.~Haissinski\inst{2}, 
J.C.~Hamilton\inst{4},
D.~Hardin\inst{1},
J.~de Kat\inst{1}, 
A.~Kim\inst{4},
T.~Lasserre\inst{1},
\'E.~Lesquoy\inst{1},
C.~Loup\inst{5},
C.~Magneville \inst{1}, 
B.~Mansoux\inst{2}, 
J.B.~Marquette\inst{5},
\'E.~Maurice\inst{3}, 
A.~Milsztajn \inst{1},  
M.~Moniez\inst{2},
N.~Palanque-Delabrouille\inst{1}, 
O.~Perdereau\inst{2},
L.~Pr\'evot\inst{3}, 
N.~Regnault\inst{2},
C.~Renault\inst{1},  
J.~Rich\inst{1}, 
M.~Spiro\inst{1},
A.~Vidal-Madjar\inst{5},
L.~Vigroux\inst{1},
S.~Zylberajch\inst{1}
\\   \indent   \indent
The EROS collaboration
}
%% 1 Saclay, 2 LAL, 3 Marseille, 4 IAP, 5 Copenhague, 6 La Serena 7 Gould
\institute{
CEA, DSM, DAPNIA,
Centre d'\'Etudes de Saclay, 91191 Gif-sur-Yvette Cedex, France
\and
Laboratoire de l'Acc\'{e}l\'{e}rateur Lin\'{e}aire,
IN2P3 CNRS, Universit\'e Paris-Sud, 91405 Orsay Cedex, France
\and
Observatoire de Marseille,
2 pl. Le Verrier, 13248 Marseille Cedex 04, France
\and
Coll\`ege de France, Physique Corpusculaire et Cosmologie, IN2P3 CNRS, 
11 pl. Marcellin Berthelot, 75231 Paris Cedex, France
\and
Institut d'Astrophysique de Paris, INSU CNRS,
98~bis Boulevard Arago, 75014 Paris, France
\and
Astronomical Observatory, Copenhagen University, Juliane Maries Vej 30, 
2100 Copenhagen, Denmark
\and
Universidad de la Serena, Facultad de Ciencias, Departamento de Fisica,
Casilla 554, La Serena, Chile
\and
Department of Astronomy, Ohio State University, Columbus, OH 43210, U.S.A.
\and
Dept. Astronom\'\i a, Universidad de Chile, Casilla 36-D, Santiago,
Chile
\and
European Southern Observatory, Casilla 19001, Santiago 19, Chile
\and
DASGAL, 77 avenue de l'Observatoire, 75014 Paris, France
}
\offprints{Nathalie.Delabrouille@cea.fr}

\date{Received;accepted}

\authorrunning{C. Afonso et al.}
\titlerunning{EROS 2 observation of SMC-98-1 caustic crossing }

\def\kms{{\rm km}\,{\rm s}^{-1}}
\def\kpc{{\rm kpc}}
\def\lsim{{\lesssim}}
\def\au{{\rm AU}}
\def\bq{{\bf ????}\ }
\def\etal{{ et al.}}

\maketitle

\begin{abstract}

	We report on intensive photometric monitoring on 18 June 1998
of MACHO SMC-98-1, a binary-lens microlensing event seen toward the
Small Magellanic Cloud (SMC).  The observations cover 5.3 hours (UT
5:17 -- 10:37), and show a sharp drop of 1.8 mag during the first 1.8
hours, followed by an abrupt flattening at UT 7:$08 \pm 0$:02.  We
interpret the kink at 7:08 as the end of the second caustic crossing
(when the source first moved completely outside the caustic).  These
results indicate that $\mu\sin\phi\lsim\,1.5\,\kms\,\kpc^{-1}$ at the
$2 \sigma$ level, where $\mu$ is
the proper motion of the lens (relative to the line of sight to the source),
and $\phi$ is the unknown (and so random) angle of the caustic crossing.
Hence, the lens probably does not lie in either the Galactic halo
or disk and so is most likely in the SMC itself.  Our data can be combined
with those of other groups to give more precise constraints on the proper
motion (and hence the nature) of the lens.
\end{abstract}

\section{Introduction}

	The EROS collaboration is engaged in long term micro\-lensing
observations toward the Large Magellanic Cloud (LMC) and Small
Magellanic Cloud (SMC) in order to determine the fraction of the Milky
Way halo in the form of massive compact halo objects.  Ten
candidate events have been detected toward the LMC, eight 
by MACHO 
(Alcock \etal\ 1997a) and two by EROS (Ansari \etal\ 1996b).  The
typical Einstein radius crossing time measured for these events is
$t_E\equiv r_E/v\sim 45\,$days, with $r_E$ the Einstein radius and $v$
the transverse speed of the lens relative to the observer-source line
of sight. The optical depth implied is of order half that required
to account for the dynamical mass of the dark halo. This time scale is
more than twice the value expected for a halo of brown dwarfs,
({\it {i.e.}}, where the mass of the objects is $M< 0.08\,M_\odot$).  Since
$r_E\propto M^{1/2}$, this would seem to imply that the lenses have
masses $M\sim 0.4\,M_\odot$. However, they cannot be main-sequence
stars since their density would then be almost two orders of magnitude
more than is observed.  Sahu (1994) and Wu (1994) have suggested that
the lenses might be in the bar/disk of the LMC itself. Dynamical
arguments seem to rule out this possibility (Gould 1995).  Numerous
other suggestions as to the nature of the lenses (Zhao 1998; Zaritsky
\& Lin 1998; Evans \etal\ 1998) have brought forth equally numerous
counter-arguments (Alcock \etal\ 1998b, Gould 1998a; Bennett 1998;
Beaulieu \& Sackett 1998).

	SMC microlensing searches provide a powerful test of the
halo-lens hypothesis.  If the lenses observed toward the LMC are
indeed in the halo, then both the optical depth and the typical
duration should be similar toward the SMC (but see Sackett \& Gould
1993).  To date, two events have been observed toward the SMC,
MACHO-97-SMC-1/EROS-SMC-1 (Alcock \etal\ 1997c; Palanque-Delabrouille 
\etal\ 1998) with $t_E \sim 123$ days and MACHO-98-SMC-1 which is
still in progress.
Both of these events are substantially longer than the average for LMC
events, but since the durations lie in the general range of the LMC
time scales, and since there are only two events, no definite
conclusion can be drawn from this comparison.

\section{Observations and Data Reduction}

	The telescope, camera, and telescope operations are as described
in Palanque-Delabrouille \etal\ (1998) and references therein.  However the  
observational strategy and data reduction differed substantially from our
previous practice.

	The event itself was electronically alerted by the MACHO
collaboration\footnote{http://darkstar.astro.washington.edu} on 25 May 1998,
just before we began a planned maintenance shutdown (26 May -- 17
June).  On 8 June, MACHO issued a secondary alert following a
dramatic increase in magnification to $A\sim 13$, indicating that the 
source had crossed a binary caustic. On 15 June, MACHO predicted a
second crossing on $19.3 \pm 1.5$ June. On 17 June, the PLANET
collaboration\footnote{http://thales.astro.rug.nl/\~{}planet/} posted photometric
data which allowed us to predict a crossing at $18.21\pm 0.08$ June,
i.e., the first night
of our resumed operations.  (PLANET independently predicted 18.0 June based
on the same data.)\ \ 
In view of the importance of caustic
crossings for understanding the nature and location of the lens (see \S\ 4), we
elected to temporarily abandon our normal monitoring strategy and to
observe only this field.  In addition, we changed our pointing so that
the lensed source would fall on a better quality CCD than during normal 
monitoring.  We conducted a continuous series of 5 min.\ exposures from 18.23 
June (when the SMC first rose above our telescope limits) until 18.35 June,
then a continuous series of 10 min.\ exposures (because the
magnification of the star had dropped significantly) until dawn at
18.45 June.

	Photometry was carried out by means of image subtraction,
using the method of (Alard \& Lupton, 1998) which we have modified,
automated, and adapted to our system.  This method differs markedly
from the point spread function (PSF) fitting program (PEIDA, Ansari 1996a)
that lies at the heart of our microlensing search.  Image
subtraction is more accurate for crowded field photometry: we
are now using PEIDA to find events but use image subtraction to measure 
their light curves.

\section{Results}

	Fig. 1 shows the flux in the EROS red and blue filters in
ADU, relative to a template constructed from 4 images from the flat
part of the curve.  Image subtraction does not yield a direct
measurement of the total flux, but no such measurement is required
for any of the analysis of this {\it Letter}.  However, to make
contact with other work, we show in Fig. 2 the position of the
source on a color-magnitude diagram at the beginning and end of the
falling part of the curve, as determined from PEIDA photometry. The
fact that the source changes color as the magnification falls shows
that it is heavily blended, although since the seeing deteriorated
rapidly during the night, it is possible that blending was a worse
problem when the fainter images were taken.  (We have done photometry
sequences on non-varying stars in the field to verify that the seeing
changes do not significantly affect the image-subtraction
photometry.)

\begin{figure} [h] 
 \begin{center} 
   \epsfig{file=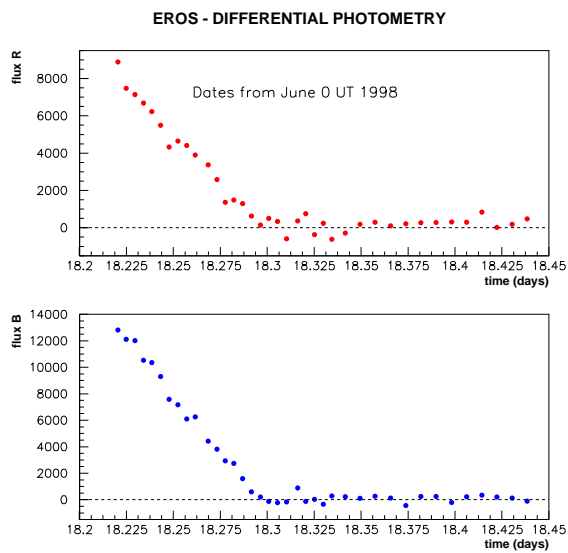,width=7.8cm} \vspace{-.1cm} 
   \caption{Differential photometry of EROS data taken on 18 June 1998. 
   R data on top, B data on bottom, in ADU.}  
 \end{center}\vspace{-0.5cm}
\end{figure}

	The blue and red curves show very similar behavior: they begin
with an almost perfectly linear decline of 1.8 mag, and then abruptly 
flatten at
UT $18.2970\pm 0.0012$ June and $18.2980\pm 0.0021$ June 
respectively.  
We interpret this break in the slope as
the end of the caustic crossing, measured accurately to within 2
min.  The ratio of the slopes of the curves (7900 ADU/hr in blue,
5700 ADU/hr in red) gives our best (i.e., blending-free) estimate of
the color of the source, $B_{\rm EROS} - R_{\rm EROS}=
-0.35^{+0.03}_{-0.04}$.  Comparison of this color with the value
at the peak ($B_{\rm EROS} - R_{\rm EROS}=-0.41$)
displayed in Fig. 2 shows that the color at the brightest PEIDA point is not
significantly affected by blending.

\begin{figure} [h] 
 \begin{center} 
   \epsfig{file=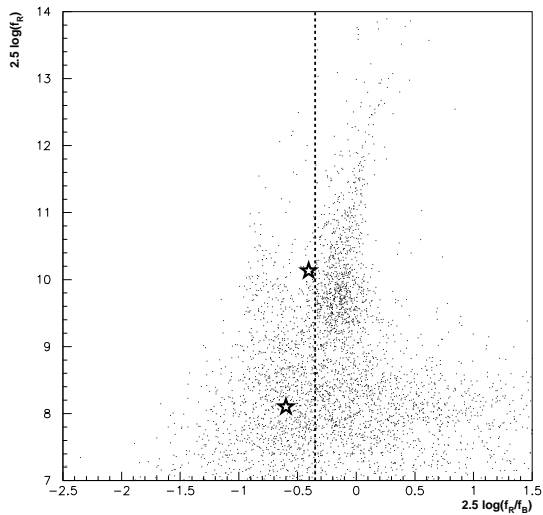,width=7.8cm} \vspace{-.1cm} 
   \caption{HR diagram of the field of the microlensing event, PEIDA photometry.
   Stars indicate the position of the source at the beginning and end of the
   falling curve. Dashed line indicates color as given by the ratio of slopes 
   in B and R. Colors are given in the EROS non standard filters.}
 \end{center}\vspace{-0.5cm}
\end{figure}

\section{Analysis}

	As we discuss below, by analyzing the complete light curve, one can
measure the proper motion of the lens relative to the observer-source
line of sight.  Since the expected proper motion of halo lenses is
$\mu_{\rm halo} \sim 220\,\kms/ 15\,\kpc\sim 15\,\kms\,\kpc^{-1}$, while that
of SMC lenses is
$\mu_{\rm SMC} \sim 30\,\kms/ 65\,\kpc\sim 0.5\,\kms\,\kpc^{-1}$
(Hatzidimitriou \etal\ 1997), one should
be able to clearly distinguish between these two possibilities.

	Unfortunately, the EROS data cover only a small portion of the light
curve.  Nevertheless, these data are sufficient to derive important
constraints.  Let $\Delta t$ be half of the total amount of time that some
part of the source is over the caustic.  The curvature of the caustic is
small compared to the source radius and so the caustic can be approximated
as a straight line.  Let $\phi$ be the angle between this line 
and the source trajectory. Then,
\begin{equation}
\mu = {\theta_*\csc\phi\over \Delta t},
\end{equation}
where $\theta_*$ is the angular radius of the source.  In principle one can 
estimate
$\theta_*$ from the color and flux of the source using the Planck law and
the (quite reasonable) assumption that the source is a black body.  
The instrumental color is
well determined from the ratio of slopes (see above).  In practice, however,
the photometry is not sufficiently well calibrated to accurately determine
the temperature from the measured color, and our pre-event data are not
of sufficiently high quality to accurately measure the unlensed flux.  We
therefore adopt a pre-event magnitude of $V=21.8$ from the original MACHO
alert.  We assume an extinction of $A_V=0.22$ 
and an SMC distance of 65~kpc 
($A_V=0.12$ foreground, from Schlegel, Finkbeiner, \& Davis (1998)
and $0.1\pm 0.1$ estimated internal extinction).
We then find $M_V=2.5$, corresponding to an A8 or F0 star with 
radius $R_*\sim 1.5 R_\odot$ (Lang 1991), and so
$\theta_*=R_*/D_{\rm SMC}\sim 0.106 \mu$as.

	Clearly, $(2\Delta t)>1.8\,$hours, since the light curve falls for
at least this length of time.  
However, we can use the smallness of the curvature of 
the falling blue light curve in Fig. 1 to place still stronger constraints on
$\Delta t$.  A binary lens gives rise to 5 images when the source is inside a 
caustic and 3 when it is outside.
As a point source approaches a caustic from the inside, 3 of the images change
only very slowly, while the remaining two diverge as $\sim (t_0 - t)^{-1/2}$,
where $t_0$ is the time of the caustic crossing.  At $t=t_0$, these two
suddenly disappear.  Hence, one can model the light curve of a point source
as
\begin{equation}
g_p(t;t_0,A,B,C) = A(t_0-t)^{-1/2} \Theta(t_0-t) + Bt + C
\end{equation}
where $\Theta$ is a Heaviside step function.  For a finite source of 
uniform surface brightness and with crossing time $2\Delta t$, the light
curve is given by
\begin{eqnarray}
\lefteqn{g(t;\Delta t, t_0,A,B,C) =} \nonumber \\ 
&& {1\over \pi(\Delta t)^2}\int_{-\Delta t}^{\Delta t} 
d s\sqrt{(\Delta t)^2 - s^2}\times g_p(t+s;t_0,A,B,C).
\end{eqnarray}

We fit the blue data (more accurate than the red data) to this form,
with errors estimated from fluctuations on the baseline and rescaled 
according to photon statistics --- this method is corroborated by
the value of the $\chi^2/dof$ 
on the fit before the kink ($14.4/15$). 
We find that fits with $\Delta t < 3$ hours
are unacceptable at the $2\,\sigma$ level ($\Delta \chi^2 > 4$) because 
they have too much curvature.  This implies
\begin{equation} \label{limitmu}
\mu\sin\phi = {\theta_*\over \Delta t}\,\lsim\;1.5\,\kms\,\kpc^{-1}.
\end{equation}

	The angle $\phi$ could be estimated from the full light curve, but
is difficult to extract from the EROS data alone.   
Better constraints could be obtained, particularly on the angle $\phi$,
if the EROS data were combined with those of MACHO, GMAN, 
and PLANET.
If the overall lensing geometry were well determined from a joint 
fit to all available data, the detailed EROS light curve of the end
of the caustic crossing would also enable us to measure the limb darkening
of the source.

\section{Discussion}

	Fig.\ 3 shows the distribution of expected values of $\mu\sin\phi$
for halo lenses
together with the upper limit from Eq. \ref{limitmu}.  If the lens is in the
halo, it sits in the extreme (7.3\%) lower end of the distribution.
The proper-motion limit is larger than typical expected SMC proper motions, 
$\mu\sim 0.5\,\kms\,\kpc^{-1}$.  However, this may be partially due to the fact
that Eq. \ref{limitmu} gives an upper limit rather than an estimate.  
If the SMC is tidally disrupted, the lens motion could also be substantially 
larger than virial estimates. At present, 
the most plausible interpretation is that the lens lies in the SMC.
\vskip -1cm 
\begin{figure} [h] 
 \begin{center} 
   \epsfig{file=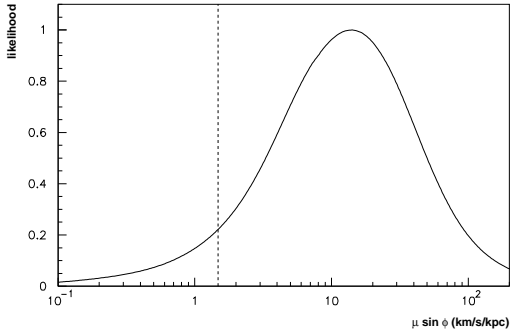,width=7.5cm} \vspace{-.1cm} 
   \caption{Distribution of expected values of $\mu\sin\phi$ for halo objects,
where $\mu$ is the proper motion and $\phi$ is the caustic-crossing angle.
Dashed line is the upper limit derived from our data.}
 \end{center}\vspace{-0.5cm}
\end{figure}

	Regarding the other SMC event, EROS-SMC-1, we placed constraints
in Palanque-Delabrouille \etal\ (1998) on
the size of the parallax effect, which can be expressed
as $R_E/(1-x) = 0\pm \sigma_{\delta u}\au$ where $\sigma_{\delta u}=0.027$.
Assuming a halo characterized by a fixed rotation speed $v_c=220\,\kms$ and
a $\rho({\bf r})\propto r^{-2}$ density distribution, one can estimate
the likelihood of the event as a function of its mass
\begin{eqnarray}
L(M) &= & M^{1/2}\int_0^1 dx [x(1-x)]^{3/2}\rho(x)\exp\biggl[-{R_E^2(x,M)\over 
(v_c t_E)^2}\biggr] \nonumber \\ 
&& \times \exp\biggl\{-{1\over 2}\biggl[{(1-x)\au\over R_E(x,M) 
\sigma_{\delta u}}\biggr]^2\biggr\},
\end{eqnarray}
where $\rho(x) \propto [1 + (xQ)^2 - 0.8 (xQ)]^{-1}$, 
$Q\sim 8.1$ is the ratio of $D_{\rm SMC}$ to the Galactocentric distance, 
$R_E^2(x,M) = 4G M D_{\rm SMC} x(1-x)/c^2$, 
and $t_E=123$ days is the Einstein
radius crossing time of the event.  This distribution is shown in Fig. 4.
Note that the peak is near $M\sim 3.2\,M_\odot$ and that 95\% of the 
distribution lies at $M>0.6\,M_\odot$.  This is highly implausible for a
halo lens unless it is a new type of object like a primordial black hole.
On the other hand, as we showed in Palanque-Delabrouille \etal\ (1998), if the
lens is in the SMC, then it is consistent with being a low mass star.

\begin{figure} [h] 
 \begin{center} 
   \epsfig{file=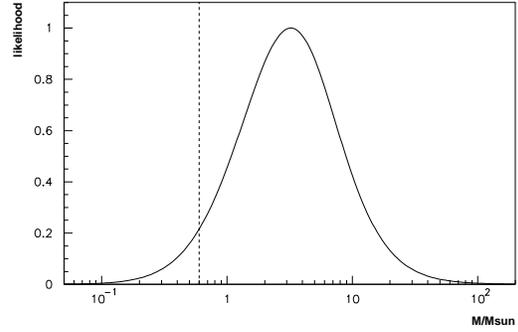,width=7.5cm} \vspace{-.1cm} 
   \caption{Mass likelihood for EROS-SMC-1 event, taking into account our
   parallax limit. 95\% of the distribution has \mbox{$M>0.6\,M_\odot$} (dashed line).}
 \end{center}\vspace{-0.5cm}
\end{figure}

Thus, both of the lenses discovered toward the SMC show significant 
evidence of being in the SMC itself.  We therefore believe that Sahu's
(1994) suggestion that the LMC events are due to self-lensing should be 
given very serious consideration notwithstanding the ``proof'' (Gould 1995)
that this idea is impossible.

	Continued (and intensified) monitoring of the SMC will be important
for testing this hypothesis.  In addition, if possible, all LMC and SMC
events should be intensively monitored for parallax effects. If the lenses
lie in the LMC or SMC, then like EROS-SMC-1, they will show no sign of 
parallax. If they are in the halo, some will show such signs (Gould 1998b).

\begin{acknowledgements}
We are grateful to D. Lacroix and the technical staff at the Observatoire de
Haute Provence and to A. Baranne for their help in refurbishing the MARLY
telescope and remounting it in La Silla. We are also grateful for the support
given to our project by the technical staff at ESO, La Silla. We thank
J.F. Lecointe for assistance with the online computing.
We wish to thank also the MACHO and PLANET collaborations, for the fruitful
exchange of information that took place during all the duration of this 
microlensing event.
\end{acknowledgements}


\begin{thebibliography}{}
\bibitem{ac} Alard, C., \&  Lupton, R.\ H.\ 1998, ApJ, submitted
\bibitem{alc97a} Alcock \etal\ 1997a, {ApJ}, 486, 697
\bibitem{alc97b} Alcock \etal\ 1997b, {ApJ}, 490, 59
\bibitem{alc97c} Alcock \etal\ 1997c, {ApJ}, 491, L11
\bibitem[Ansari 1996a]{Peida} Ansari R. (EROS coll.), 1996a, 
  {Vistas in Astronomy} {40}, 519.
\bibitem[Ansari 1996]{ans96} Ansari R. (EROS coll.), 1996b, 
  {A\&A} {314}, 94.
\bibitem{BS} Beaulieu, J.-P., \& Sackett, P.\ D.\ 1998, {AJ}, submitted 
(astro-ph 9710156)
\bibitem{alc98} Bennett, D.\ 1998, {ApJ}, 493, L79
\bibitem{evans} Evans, N.\ W., Gyuk, G., Turner, M.\ S., \& Binney, 
J.\ J.\ 1998, 
Nature, submitted
\bibitem{gthree} Gould, A.\ 1995, {ApJ}, 441, 77
\bibitem{gtwo1} Gould, A.\ 1998a, {ApJ}, 499, 728
\bibitem{gtwo2} Gould, A.\ 1998b, {ApJ}, 506, 000
\bibitem[Hatzidimitriou \etal 1997]{Hatzi2} Hatzidimitriou D. \etal, 1997, 
  {AA Suppl.} 122, 507.
\bibitem{lang} Lang, K.R.\ 1991, Astrophysical Data: Planets and 
Stars (New York:Springer-Verlag)
\bibitem{pb} Palanque-Delabrouille, N.\ \etal\ 1998, {A\&A}, 332, 1
\bibitem{sg} Sackett, P.\ D.\ \& Gould, A.\ 1993, ApJ, 419, 648
\bibitem{sahua} Sahu, K.\ C.\ 1994a, Nature, 370, 275
\bibitem[Schlegel, Finkbeiner, \& Davis 1998]{schlegel} 
Schlegel, D.\ J., Finkbeiner, D.\ P., \& Davis, M.\ 1998, {ApJ}, 500, 525
\bibitem{wu} Wu, X.-P.\ 1994, {ApJ}, 435, 66
\bibitem{zarit} Zaritsky, D., \& Lin, D.\ N.\ C. 1997, {AJ}, 114, 254
\bibitem{zhao} Zhao, H.\ 1998, {MNRAS}, 294, 139
\end{thebibliography}
\end{document}